\documentclass[prb,twocolumn,groupedaddress,amssymb]{revtex4}
\usepackage{epsfig}

\usepackage{bm}

\def\be{\begin{equation}}
\def\ee{\end{equation}}
\def\ba{\begin{eqnarray}}
\def\ea{\end{eqnarray}}

\begin{document}
\title{On the magnetization of two-dimensional superconductors}
\author{Vadim Oganesyan}
\author{David A. Huse}
\author{S. L. Sondhi}
\affiliation{Department of Physics, Princeton University,
Princeton, NJ 08544, USA}

\date{\today}

\begin{abstract}
We calculate the magnetization of a two-dimensional superconductor
in a perpendicular magnetic field near its Kosterlitz-Thouless
transition and at lower temperatures.  We find that the critical
behavior is more complex than assumed in the literature and that,
in particular, the critical magnetization is {\it not} field
independent as naive scaling predicts.  In the low temperature
phase we find a substantial fluctuation renormalization of the
mean-field result.  We compare our analysis with the data on the
cuprates.
\end{abstract}
\pacs{PACS numbers: xxx, xxx, xxx} \maketitle

\section{Introduction}

The study of fluctuation superconductivity received a tremendous
impetus from the discovery of the cuprate, high temperature
superconductors\cite{BlatterRMP}. It was quickly realized that
these were materials with large Ginzburg parameters with the
additional feature that many of them are also highly
two-dimensional. More recently, interest has focussed on the
systematics of their superconducting properties with doping and it
has been found that this anisotropy is further enhanced with
underdoping\cite{Schneider}. Conversely, the extent to which the
anomalous properties of the underdoped compounds in the
``pseudogap'' region can be attributed to superconducting
fluctuations is a question of considerable
interest\cite{OrensteinMillis,WangM,WangN,PseudoOng,IU,EK,Mohit}.

In this context, we report here a study of the magnetization of a
two-dimensional superconductor, or more precisely of a stack of
decoupled two dimensional superconducting layers, in a
perpendicular magnetic field $H$.
By combining Kosterlitz-Thouless renormalization group flows and
explicit computations for plasmas, we find the field and
temperature dependence of the magnetization density, $M(H, T)$,
for temperatures $T$ near to or below the Kosterlitz-Thouless
transition temperature $T_{KT}$ at fields $H_{c1}\ll H\ll H_{c2}$.
These results are
interesting on three immediate fronts. First, the rough magnitude
of $M(H, T)$ and the trends exhibited in Fig. 1 are in accord with
much of the data on the strongly layered cuprates in the field
regime where a single layer theory is expected to apply. For
example, this holds for the recent data of Wang and
collaborators\cite{WangM} which have demonstrated the existence of
substantial fluctuation magnetization in the same region of the
phase diagram of BSCCO which was previously observed to exhibit a
large Nernst signal\cite{WangN}. Second, we find that $M$ is {\it
not} independent of $H$ at criticality, as assumed in earlier
analysis of the ``crossing point'' phenomenon in studies of the
cuprates\cite{crossing}. Instead, the critical magnetization
follows \ba M \approx -\frac{k_B T_{KT}}{d \phi_0}\log \gamma_1
\log\frac{\phi_0}{\mu_0 Ha_0^2 \gamma_2} \label{eq:critmag} \ea
where $d$ is the distance between layers, $\phi_0$ is the flux
quantum,
$H$ is the magnetic
field component perpendicular to the layers and $a_0$ is the
microscopic short-distance cutoff length ($\gamma_1$ and
$\gamma_2$ are constants). Third, we find that below $T_{KT}$, $M$
exhibits a low field growth indicative of the expulsion of
vortices from the KT phase,
\ba
M \approx - \frac{\pi \rho_s(T)}{ d \phi_0}\left(
1-\frac{2 k_B T}{\pi \rho_s(T)}\right)\log \frac{\phi_0}{\mu_0 H a_0^2
\gamma_3}, \label{eq:lowT} \ea where $\rho_s$ is the 2D superfluid
density, related to the in-plane penetration depth via $\rho_s= d
\phi_0^2/(4\pi^2\lambda^2\mu_0)$
($\gamma_3$ is a
constant that vanishes as $T$ approaches $T_{KT}$, see below).
Note that this expression gives a simple relation between the
superfluid density and the derivative of the magnetization with
respect to the logarithm of the field, that can be
used to determine when the system is indeed behaving as a stack of
essentially decoupled two-dimensional Kosterlitz-Thouless films.

These results and the limits of their applicability to actual
layered superconductors are derived in Section II, with some
details relegated to an Appendix. In Section III we consider
experimental signatures of our analysis and the existing
experimental situation. We close with a brief summary and some
open questions.

\section{Theory}
\label{sec:theory}

\subsection{Definition of the problem}
The average magnetization density of a superconductor in an
external field ${\bf H}$ is

\ba
\mu_0{\bf M}={\bf B}({\bf H})-\mu_0 {\bf H},
\ea where
$\bf{B}(\bf{H})$ is the average (uniform) magnetic field,
computed via
\ba
\frac{\partial f_{scm}(\bf{B})}{\partial
\bf{B}}=\bf{H},
\ea
where $f_{scm}({\bf B})$ is the total free energy
density of superconducting matter coupled to fluctuating magnetic
fields with average field fixed at ${\bf B}$. Let us define
\ba
&&f_{sc}=f_{scm}-\frac{{\bf B}^2}{2\mu_0}, \ \ {\rm then}\\
&&\frac{\partial f_{sc}}{\partial {\bf B}}=\frac{\partial
  f_{scm}(\bf{B})}{\partial \bf{B}}-\frac{\bf B}{\mu_0}=-{\bf M}.
\ea In type II materials and for $H\gg H_{c1}$ field energy is
well-approximated by its uniform value, and therefore, following
the equation above, the magnetization is to be computed from the
free energy of a charged superfluid in a uniform \emph{external}
magnetic field.

We consider a stack of decoupled two dimensional layers. Hence we
will replace the three dimensional free energy density
$f_{sc}({\bf B})$ by ${1 \over d} f (B)$ where $f(B)$ is the free
energy density of a single layer in the presence of a magnetic
induction $B$, now restricted to be perpendicular to the layers, whose effect will be
to impose a density
$B/\phi_0$ of field-induced vortices.

\subsection{Coulomb gas formulation}
\label{sec:coulomb} To compute $f$ and thence $M$ we shall resort
to the standard\cite{Minnhagen} mapping of the two-dimensional vortex problem onto
a two component Coulomb plasma whose Hamiltonian is given by (see
Appendix A)
\ba
H=N_T E_{c0} + \pi \rho_{s0} \sum_{i<j} p_i p_j \log
\frac{r_{ij}^2}{a_0^2}+H_{\rm B}.
\ea The number of vortices of
charge $p_i = \pm 1$ is $N_{\pm}$. The total number of vortices is
$N_{\rm T}=N_+ + N_-$ and is allowed to fluctuate by the addition
and removal of neutral vortex-antivortex pairs, but the net charge
$Q=N_+ - N_-=L^2 B/\phi_0$ is constrained by the field $B$.  $E_{c0}$ is
the bare vortex core energy, $a_0$ --- the bare short distance cutoff
(e.g., vortex core radius), $\rho_{s0}$ is the bare superfluid
stiffness (inverse dielectric constant in the plasma language),
$L^2$ is the area of the system.  A uniform background density of
charge to make the system neutral is necessary to insure a proper
thermodynamic limit.  The total Coulomb interaction potential
between the vortices and this background and between the
background and itself is in the constant $H_{B}$ that depends on
$B$ and on the size and shape of the system, but not on the vortex
configuration.

\subsection{Renormalization and matching}
Standard Kosterlitz renormalization group (RG)
methods\cite{Kosterlitz,AndersonYuval,Minnhagen} can be
generalized to the present, non-neutral situation. Our strategy
will be to renormalize until we can match on to a solution by an
exact or approximate method.
For convenience, the bare
superfluid stiffness and core energy are captured by the bare
couplings $x_0=1-\frac{\pi \rho_{s0}}{2 k_B
  T}$ and $y_0=2\pi e^{-E_{c0}/k_B T}$, which we will renormalize.
In addition to these couplings we shall be interested in computing
the free energy density, so we keep track of the
configuration-independent term $\mathcal{C}$ in the renormalized
Hamiltonian, generated as degrees of freedom are integrated out,
i.e., at any intermediate step in the RG process the partition
function is $Z=e^{-L^2 \mathcal{C}/a^2} {\rm Tr} e^{-\beta H}$,
where $a$ is the renormalized cutoff.  Note $\mathcal{C}$ is
defined to be dimensionless.  We also define the dimensionless density
of field-induced vortices (number per
renormalized cutoff area) as $n=B a^2/\phi_0$; its bare value is $n_0=B a_0^2/\phi_0$.

Provided the total number of vortices and antivortices per cutoff
area is small compared to one, the following differential
equations describe the renormalization upon increasing the cutoff
to $a= a_0 b$ and integrating out neutral vortex-antivortex pairs
with spacing less than $a$: \ba \frac{d y}{d \log b}&=&2xy
\label{eq:dy}\\
\frac{d x}{d \log b}&=& 2y^2
\label{eq:dx}\\
\frac{d \mathcal{C}}{d \log b}&=&2 \mathcal{C} - \frac{y^2}{2\pi}
\label{eq:dA}\\
\frac{d n}{d \log b}&=& 2 n \ \ . \ea The last line shows the
trivial renormalization of the number of field induced vortices
per cutoff area.
A somewhat unexpected fact about these equations is that the
presence of field induced vortices leaves the zero-field flow
equations (Eqs.~(\ref{eq:dy}), (\ref{eq:dx}) and (\ref{eq:dA}))
unaffected (also see Sec. IV). As we shall see, this simplifies
the calculation of the magnetization.

By straightforward integration one obtains, with
$$c\equiv|x_0^2-y_0^2|^{1/2}\sim\sqrt{|T-T_{KT}|},$$
the solutions
\ba
\label{eq:xbelow}
x_{<}(b) &=&c\ \rm{cotanh} (\Theta_<+2c \log b)\nonumber \\
y_{<}(b) &=&c\ \rm{cosech} (\Theta_<+2c \log b) \nonumber \\
\mathcal{C} (b) &=& b^{2}\mathcal{C} (0) - b^2 \int_0^{\log
b}\frac{y_{<}^2 (b')}{2 \pi b'^2 } d \log b' \nonumber \\
n (b) &=& n_0 b^2
\ea
with $\cosh \Theta_<=-\frac{x_0}{y_0}$ for $T < T_{KT}$, and the
solutions,
\ba
x_{>}(b)&=&c\ \rm{cotan} (\Theta_>-2c \log b) \nonumber \\
y_{>}(b) &=&c\ \rm{cosec} (\Theta_>-2c \log b)  \nonumber \\
\mathcal{C} (b) &=& b^{2}\mathcal{C} (0) - b^2 \int_0^{\log
b}\frac{y_{>}^2 (b')}{2 \pi b'^2 } d \log b' \nonumber \\
n (b) &=& n_0 b^2 \nonumber \\
\ea
with $\cos \Theta_>=\frac{x_0}{y_0}$ for $T > T_{KT}$.

The free energy density of the original problem is then computed
by running the RG to scale $b_R$, where it can be expressed as a
sum of two parts \ba f=\frac{\mathcal{F}_R+k_B T\mathcal{C}(b_R)}{a_R^2}, \ea
where $\mathcal{F}_R$ is the free energy per renormalized cutoff area of the vortices that have
not yet been integrated out.  [Note: since we have chosen
to rescale lengths in our RG, the free energy density $f$ is computed by dividing the residual free energy, $\mathcal{F}_R$, and $\mathcal{C}(b_R)$, by the rescaled area, $a_R^2$.]  The observation that field induced vortices do not affect
the other flows, in particular that $\mathcal{C}$ has no explicit
dependence on $B$, simplifies the calculation of the
magnetization density: \ba \label{eq:magnetiz} M=- \frac{1}{d} \frac{\partial f}{\partial B}= -
\frac{1}{d \phi_0}
\frac{\partial \mathcal{F}_R}{\partial n_R}, \ea thus reducing the
problem to computing $\mathcal{F}_R$.

The renormalized density of field-induced vortices, $n$,
grows
under the RG flow.  For $T \leq T_{KT}$ thermally induced vortices
become increasingly dilute and the system approaches a one
component plasma in the low-field limit where the flow can
continue to large $b_R$. For $T
> T_{KT}$, the renormalized density of thermal vortices first decreases and
then increases, so that the system is always a two-component
plasma with a scale-dependent charge imbalance. Our principal
results for the asymptotic behavior of magnetization (Eqs. (1) and
(2)) follow from the exactly known free energy of a dilute one
component plasma\cite{Jancovici} whose parameters, the charge
density, core energy
and dielectric constant, are given by the RG flows above.  A
Debye-H\"uckel approximation for the free energy of the
two-component plasma reproduces the correct low field behavior for
$T\leq T_{KT}$ and provides a reasonable approximation elsewhere,
including $T > T_{KT}$; we shall use it to generate plots of the
magnetization versus field in the vicinity of the transition.
\subsection{Results}
\subsubsection{Low fields, $T\leq T_{KT}$}
For $T\leq T_{KT}$ the above Kosterlitz-Thouless RG flows go to
small fugacity $y$, where the leading terms in the exact free
energy are known for low density (small field).  Thus we run the
RG to a matching scale $a_R=b_R a_0$, where the parameters are
$n_R=Ba_R^2/\phi_0$,
$x_R$ and $y_R$.  We stop and match when $n$, which is growing,
becomes of order $y$.  The free energy density of the remaining
vortices, to leading order in the small parameters $n_R$ and
$y_R$ is\cite{Jancovici} \ba \mathcal{F}_R=n_R k_B
T(\log\frac{2\pi}{y_R}+ x_R \log n_R +{\cal O}(1))~.
\label{eq:ocp} \ea  The first term is the (renormalized) core
energy of the field-induced vortices, while the second term
contains their entropy and (renormalized) interaction energy.  The
thermally-excited vortex-antivortex pairs do not enter at this
order.  Thus, from (\ref{eq:magnetiz}) the magnetization is
\ba\label{eq:m} M = -\frac{k_B T}{d
\phi_0}(\log\frac{2\pi}{y_R}+x_R \log n_R+{\cal O}(1))~.
\label{eq:mocp} \ea

To start, we consider the regime of small fields below $T_{KT}$.
For $b_R^2\gg e^{2\gamma_4/c}$, where $\gamma_4$ is a constant, the renormalized parameters are computed
from the Kosterlitz RG equations as (cf. Eq.~(\ref{eq:xbelow})) \ba
x_R&\rightarrow&-c\sim-\sqrt{T_{KT}-T}\\
y_R&\rightarrow&\frac{2c y_0}{b_R^{2c}(c+|x_0|)}, \ea so the low
field magnetization is \ba M=-\frac{k_B T}{d
\phi_0}(c\log\frac{\phi_0}{B a_0^2} +
\log\frac{\pi(c+|x_0|)}{cy_0}+{\cal O}(1))~, \ea
in agreement with our Eq.~(2) above ($\gamma_3 \sim c^{1/c}$). The
low field regime of validity of this expression requires $n_R\ll
1$, which, with the above constraint on $b_R$, translates to \ba
B\ll\frac{\phi_0}{a_0^2}e^{-2\gamma_4/c}~. \ea Observe that this
condition defines a crossover length scale \ba \xi_< \sim a_0
e^{\gamma_4/c}~, \ea which has the functional form of the
correlation length above the transition, and yet is defined---like
the Josephson correlation length in Goldstone phases---{\it below}
the transition.

For higher fields and
temperatures near $T_{KT}$, there is a crossover to the critical
behavior (1), that we derive below.  Note that the slope of $M$
vs. $\log B$ at low field vanishes linearly in $c$ as $T$
approaches $T_{KT}$ from below, while the crossover field below
which this is the behavior vanishes exponentially.

If, on the other hand, one is near the Kosterlitz-Thouless
transition (so $c\ll 1$), the RG flows for $1\ll b_R^2\ll
e^{2\gamma_4/c}$ follow \ba -x\approx y\approx\frac{1}{2 \log b}~.\ea Here
when we match at $n_R\approx y_R$, it is at
$y_R\approx-\frac{1}{\log (B a_0^2/\phi_0)}$ and the resulting
magnetization is
\ba M= -\frac{k_B T}{d \phi_0}(\log\log \frac{\phi_0}{B a_0^2}
+{\cal O}(1))~, \ea which is the same as Eq.~(1).  The field range
where this result applies is \ba \frac{\phi_0}{a_0^2}\gg
B\gg\frac{\phi_0}{\xi_{<}^2}~, \ea which is an intermediate field
range for $T$ near to but below $T_{KT}$, and is all low fields
for $T=T_{KT}$.  Our method of analysis in principle also gives
the functional form of the crossover scaling function between
these two regimes, but we have not been able to write this
function in any concise form.  However, below we obtain an
approximate scaling function covering both regimes.

It is worth noting that in the vicinity of $T_{KT}$ the matching,
which we choose to do at $n_R\cong y_R$, occurs at $|x_R|\ll
1$, so it is the renormalized core energy (the first term in
Eq.~(\ref{eq:mocp})) that is dominant in determining the magnetization.

\subsubsection{Approximate results, all $T$}
Building on this last observation we now consider a Debye-H\"uckel
mean field theory whereby the matching free energy is approximated
by the contributions from the renormalized core energy and the
entropy. This will allow us to obtain numerical results for the
magnetization which are consistent with our previous exact results
and can be extended to above $T_{KT}$.  However, we must caution
that these are approximate results, and the precise quantitative
values of the magnetization do depend on aspects of the
calculation that are not {\it a priori} constrained by what we
already know about these materials.  These unconstrained freedoms
in the approximation we present below include ($i$) the form of
the matching free energy, in particular one might also include an
approximation to the energy of vortex-vortex interactions, ($ii$)
the scale at which the matching is done, and ($iii$) the bare
vortex core energy.

\begin{figure}
\centering
\epsfig{file=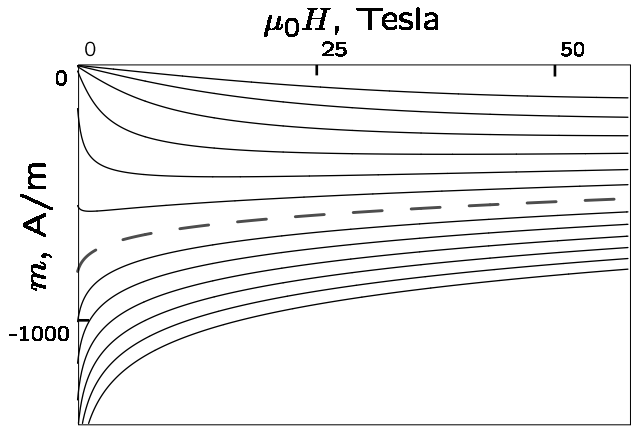,width=3.75in,bb=50 250 790 730 }
\caption{Field dependence of magnetization for a range of
temperatures between $T=77 K$ and $T=83K$, with the dashed
 curve corresponding to $T=T_{KT}=80K$ (others are spaced in $0.5 K$ increments). Here we take $E_c=\pi
\rho_{s0}(T)$, which then implies $k_B T_{KT}\approx 1.13 \rho_{s0}(T_{KT})$.
We chose $a_0=30 A$, $d=15 A$, so that
$B a_0^2/\phi_0 \approx 0.04$ at 10 Tesla and $k_B T_{KT} /(d \phi_0)\approx 350 A/m$.
Finally, we model the experimentally observed temperature dependence of the bare superfluid stiffness by
$\rho_{s0}(T)/\rho_{s0}(0)=1-T/120K$
(see, e.g., Ref.~\onlinecite{morgan}).
} \label{fig:DH}
\end{figure}

\begin{figure}
\centering
\epsfig{file=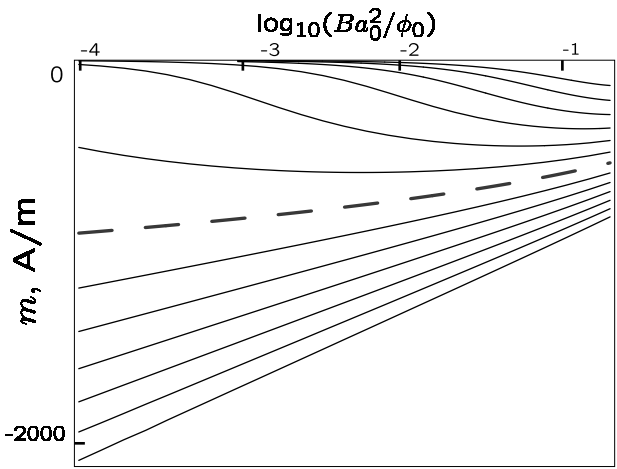,width=3.75in,bb=50 280 790 730 }
\caption{Semi-logarithmic plot of magnetization. Parameters are as
in Fig. 1.} \label{fig:DHLog}
\end{figure}

Specifically, we approximate the residual free energy density at
matching as \ba \mathcal{F}_R=(n_R^{+} + n_R^-)E_{cR}
 + k_B T (n_R^+ \log n_R^+ + n_R^- \log n_R^-)~,
\label{eq:dhF} \ea where vortices and
antivortices remaining at matching are at densities $n_R^+$ and $n_R^-$,
respectively.  The field constrains the difference between these
densities to be $n_R^{+}-n_R^{-}=n_R$, but their mean,
$\overline{n_R}=\frac{n_R^+ + n_R^-}{2}$, can vary and will take on the value
that minimizes the free energy:
\ba
\frac{\partial
\mathcal{F}_R}{\partial \overline{n_R}}= 2 E_{cR}+ k_B T \log
   [\overline{n_R}^2-\frac{n_R^2}{4}] + 2k_B T = 0 \ .
\ea
The matching condition is defined by asking that the average
density of vortices be $1$:
\ba
\overline{n_R}=\sqrt{\frac{n_R^2}{4}
+\left(\frac{y_R}{2\pi e}\right)^2}\equiv 1
     \label{eq:b}
\ea and this is used to solve for $b_R(B, T)$. The choice of
precisely unit density is clearly arbitrary---we make it for
concreteness. 
With these
assumptions the magnetization is given by  \ba M&&=-\frac{k_B
T}{2d\phi_0}\log \frac{1+ \frac{n_R}{2}}{1 -
  \frac{n_R}{2}}\\
&&=-\frac{k_B T}{d\phi_0}\log \frac{1+ \sqrt{1-(y_R/2\pi
e)^2}}{y_R/(2\pi e)}
\label{eq:DH}
\ea
which is plotted in Figs. \ref{fig:DH}, \ref{fig:DHLog} and \ref{fig:DHnonX}.

Above  $T_{KT}$ the RG flows at asymptotically small fields
are terminated at a finite fugacity and $a_0 b \sim \xi$
(the zero field correlation length), so that as $B\rightarrow 0$
the magnetization vanishes as
\ba
M\approx -\frac{k_B T}{2d\phi_0^2}\xi^2 B
\equiv -\frac{k_B T a_0^2}{2d\phi_0^2}
e^{\pi/c}  B .
\ea
This expression is
consistent with an earlier linear response
result\cite{HalperinNelson}.

\subsubsection{Comments}
Finally, three comments are in order. \\

First, the divergence of $M$ below $T_{KT}$ signals the expulsion
of vortices from that phase---indeed the logarithmic divergence of
the free energy of a single vortex with system size. The
divergence we find {\it at} $T_{KT}$ likewise signals the now
weaker expulsion of vorticity; At $T_{KT}$ the free energy of a
single vortex grows with system size as $\log(\log L)$.

Second, the divergence of $M(B)$ at low $B$ is a correct statement
about the {\it magnetic induction} dependence of the free energy
of a single layer in the limit of infinite penetration depth.  It
does not imply the divergence of an actual three dimensional
magnetization density.  For a stack of planes, 
the self-consistency implicit in Eq.~(6) keeps $M(\mu_0 H)$ from
blowing up---instead we get the Meissner phase ($B=0$) at
sufficiently low applied fields.  
Our results are valid in the experimentally relevant regime $M \ll
H$ (equivalently $H\gg H_{c1}$) where this distinction is not
important.

Finally, for $T{\not =}T_{KT}$ as we have already remarked
$B^*\approx \frac{\phi_0}{\xi_<^2}$ separates the small field
behavior from the critical $\log \log 1/B$ dependence at larger
fields. Interestingly, the magnetization displays a noticeable
temperature dependence even at these larger fields. It is
noteworthy that this dependence is opposite to that implied by the
explicit prefactor of temperature (e.g., in Eq.~(\ref{eq:DH})),
which by itself predicts an unintuitive enhanced diamagnetism at
higher temperatures. To understand this important detail consider
raising $T$ in the vicinity of $T_{KT}$. This leads to an increase
in $y_R$, which in turn gets strongly amplified by the singularity
at $y_R=0$ of the functional dependence $M\sim\log{
1/y_R}$. This variation always overwhelms the contribution from
the prefactor of $k_BT$,
so the combined effect is a reduction of $|M|$ upon raising the
temperature, as expected.

\subsection{Dimensional crossovers in layered and quantum systems}
\label{sec:layered} In sufficiently anisotropic materials we
expect that various weak interlayer couplings present will only
affect the physics at long length scales and can therefore be
treated to simply restrict the domain of applicability of the
purely two-dimensional treatment above.  As the application of the
magnetic field introduces a new length scale, the intervortex
separation or the magnetic length, this restriction means our
two-dimensional behavior will cross over to three-dimensional
behavior near a low crossover field.  There are two basic types of
interlayer interactions here: Josephson couplings induced by
charge fluctuations transverse to the layers, and electromagnetic
couplings between currents in different layers mediated by the
fluctuating magnetic field.  We consider their effects in turn.

The Josephson coupling, $J$, will a) induce true three-dimensional
long range order in the low temperature phase in zero field and
alter the universality class of the transition to that of the 3DXY
model, b) shift the transition temperature upwards to $T_c$ and c)
cause the vortices to crystallize three dimensionally at any
temperature below $T_c$ for fields less than the melting field
$B_m(T)$. While this is a problem that has been studied in some
detail \cite{KoshelevMC} we can get a feeling for the scales
involved by considering renormalization of $Ja^2$, the Josephson
coupling per cutoff area, near the (decoupled layer)
Kosterlitz-Thouless fixed point, \ba \frac{
\partial (Ja^2)}{\partial \log b}=(2-\frac{k_B T}{2\pi \rho_s})(Ja^2).
\ea From this equation we can obtain the scaling of both the shift
of the critical temperature in zero field, $\delta T_c$, and the
field scale, $B_J$, marking the crossover between two- and
three-dimensional behaviors. This is done by evaluating the length
scale at which the renormalized value of Josephson coupling is
comparable to the in-plane stiffness, $Ja^2\approx \rho_s$ (which
we shall set to its critical value $\rho_s=2k_B T_{KT}/\pi$, since
its renormalization is comparatively less important than that of
$Ja^2$). The upward shift of the critical temperature due to the
Josephson coupling, $\delta T_c$, can be estimated by identifying
this scale with the zero-field correlation length, giving \ba
\frac{\delta T_c}{T_{KT}}=\frac{(7\pi \gamma_5/8)^2}{\log^2
\frac{k_B T_{KT}}{\gamma_6 J_0a_0^2}}, \label{eq:Tc} \ea where
$\gamma_5$ is a non-universal constant related to the behavior of
$c$ near $T_{KT}$ via $\gamma_5 c\approx\sqrt{(T-T_{KT})/T_{KT}}$,
$\gamma_6$ is another constant of order one, and $J_0$ is the bare
Josephson coupling per unit area. Similarly, by substituting the
magnetic length in place of $\xi$, we find for the crossover field
at $T_{KT}$,
\ba \label{eq:Bstar}
B_J=\frac{\phi_0}{a_0^2}\left(\frac{J_0a_0^2}{k_B T_{KT}}\right
)^{8/7}. \ea

The case of electromagnetic coupling alone is a little muddier.
The situation in finite magnetic fields is that the vortices in
different layers now experience an attraction and thus can
crystallize three dimensionally even when a single layer is a
vortex liquid.  Naively, the crossover field is of order the
melting field, which is estimated to be
\cite{BlatterRMP,Clem} \ba
B_B &\approx& \phi_0/\lambda^2 \nonumber \\
    & \sim & \frac{\phi_0}{a_0^2}\left(\frac{\mu_0 k_B T_{KT}a_0^2}{
\phi_0^2 d} \right) ~,\ea where $\lambda$ is the penetration
length at $T_{KT}$.  The impact of these magnetic interlayer
couplings on the zero field transition does not appear to be a
settled problem.
The interlayer interactions are also logarithmic and thus are
marginal operators at face value. This has led to assertions that
the actual transition is still in the Kosterlitz-Thouless
universality class \cite{BlatterRMP}. However, a renormalization
group analysis by Timm \cite{Timm} finds flows at variance with
this scenario.  For our present purposes, we note that this
crossover field is, for the case of large $\kappa$ that we are
interested in, less than $\mu_0H_{c1}$; it is this crossover that
generally restricts the validity of our two-dimensional analysis
to $B\gg\mu_0H_{c1}$.

The scales $B_J$ and $B_B$ mark the rough boundary between two and
three dimensional physics.  At $T_{KT}$ the low-field three
dimensional state is crystalline, and the high-field two
dimensional dependence of $M$ crosses over to fairly standard (but
with renormalized parameters) low-field behavior in
the Abrikosov vortex lattice.
At higher temperature at the three dimensional $T_c$, in the
presence of Josephson couplings, the crossover is instead to the
$M\sim -\sqrt{H}$ behavior expected at low field in the 3DXY
critical regime.

Finally, one additional crossover, this time at large fields, is
possible when the thermal phase transitions take place in
proximity to a quantum phase transition out of the superconducting
state---as may be germane in the case of the underdoped cuprates.
Standard scaling, when applied to magnetization of the
$2+1$-dimensional quantum critical theory predicts $M\sim
-\sqrt{H}$, with crossover to this behavior taking place at
sufficiently large fields where quantum fluctuations of the order
parameter are important. The transition between the forms derived
in this paper and this regime would be a striking signature of
such fluctuations.

\subsection{Prior work}
\label{sec:prior} Early experiments on the most anisotropic
cuprates saw a near crossing point for the curves $M(T)$ taken at
varying fields $B$. This is equivalent to the statement that they
found a temperature at which the magnetization was field
independent. This was interpreted in the scaling framework
\cite{Schneider} as evidence for two dimensional critical behavior
with the one-parameter scaling form, \be f(t,B) = {1 \over \xi^2}
\tilde{f}(B \xi^2) \ee for the free energy density. As we have
shown in this paper, the scaling is not so simple at the
Kosterlitz-Thouless transition, with its two marginal operators,
and magnetization instead has a weak double-logarithmic dependence
on the field. Aside from this general scaling argument, there are
three types of prior computations that we are aware of (not listed
in chronological order):

First, Gaussian fluctuations for the Ginzburg-Landau theory in
$d=2$, yield a field independent magnetization at the Gaussian
(mean field) transition temperature\cite{Koshelev}.
This is consistent with the
absence of any marginal operators at this unstable fixed point.
Our calculation replaces this result as an account of the true
low-field critical scaling of $M$.

Second, the so called ``lowest Landau level" approximation has
been used by Tesanovic and others to study strong amplitude
fluctuations at high fields near mean-field $H_{c2}$ (see, e.g.,
Ref. \onlinecite{ZBT}). Calculations of this kind have no
overlapping regime of validity with ones presented above for the
low field regime. It would be worthwhile to see if the two sets of
results can together capture the behavior over the entire range of
fields of interest.

Finally, Bulaevskii, {\it et al.}\cite{Bulaevskii} have considered
the effects of thermal fluctuations on the magnetization of an
Abrikosov lattice in a layered superconductor with the Josephson
coupling being the dominant source of the three-dimensional order.
Treating the phonons of the lattice to quadratic order they
compute the entropic correction to the Ginzburg-Landau-Abrikosov
free energy. At low fields $B \ll B_{cr}$
they find a field independent correction to the leading
(Abrikosov) logarithm. For $B \gg B_{cr}$ they report a correction
which is itself a logarithm and leads to an expression identical
to our Eq.~(2). The scale $B_{cr}$ is the Lindemann estimate for
the melting field of the vortex lattice they begin with, so the
computation is not really valid in the regime of the logarithmic
dependence where the system is now a vortex liquid. Nevertheless,
at sufficiently low fields $B\ll\phi_0/\xi_<^2$ and $T< T_{KT}$
their result and ours agree, a sign that the vortex liquid is
locally quite similar to the crystal for these parameters. At
higher fields and at $T_{KT}$, however, their reasoning breaks down
as our calculation explicitly demonstrates.

\section{Experiments}
We now turn to the existing data on the cuprates, the systems that
have motivated this work, and some suggestions for experimental
tests of the theory.

Many of the cuprates are highly two dimensional and appear to
become increasingly so with underdoping.  The two most commonly
used diagnostics of anisotropy are the resistivity and superfluid
density tensors. Using either of these in materials such as
BSSCO-2212 we arrive at estimates of the anisotropy of order
$10^{-4}$ to $10^{-6}$ between the ab (Cu-O) planes and the
c-axis. A somewhat more direct measure of the anisotropy can be
obtained from observations of the c-axis plasma resonance due to
interlayer Josephson coupling. From the results of Ref.
\onlinecite{OngJ} for Josephson coupling per area $J_0 \sim
10^{-8} {\rm Joule/m^2}$ we obtain \ba \frac{J_0 a_0^2}{k_B
T_c}\approx 10^{-6}. \ea For such anisotropies and other
parameters appropriate to the cuprates, e.g., as in caption of
Fig. 1, we find $\delta T_c/T_{KT} \approx 0.02$, $B_J \approx
0.0002$ Tesla and $B_B \approx 0.004$ Tesla using the estimates derived
in Section  IIE. This indicates that our two dimensional theory
should give a useful account of the superconducting fluctuations
for a reasonable range of fields.

The scale for magnetization effects is set by $\frac{k_B
T_{KT}}{d\ \phi_0}$ which is $\approx 350A/m$ for our parameters.
The dimensionless factors multiplying it in our expressions are
not too different from unity.  This order of magnitude estimate is
consistent with experimental observations
\cite{WangM,Martinez,Kogan}.  The trends in the temperature and
field dependence of the magnetization we show in our figures here
are mostly in good qualitative agreement with those seen in the
experiments \cite{WangM,Martinez,Kogan}.
The most detailed published investigation is that by Kogan and
collaborators\cite{Kogan} (see also Ref.\ \onlinecite{Martinez})
who were inspired by the predictions of Ref.\
\onlinecite{Bulaevskii}.  They reported evidence that $M \sim \log
H$ with a coefficient that changes sign at  $T_{KT}$.  While this
claim below  $T_{KT}$ is consistent with our analysis, this is not
so above  $T_{KT}$ where the functional form is no longer a
logarithm.  This suggests that a reexamination of that regime is
in order.

\begin{figure}
\centering
\epsfig{file=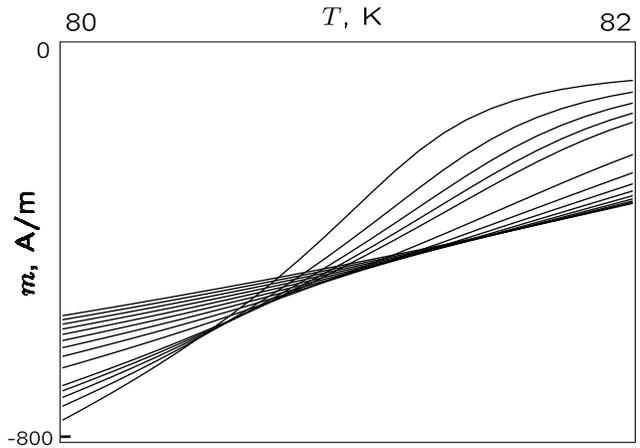,width=3.75in,bb=50 220 790 730 }
\caption{Magnetization at 1, 2, 3, 4, 5, 10, 15, 20, 25, 30, 35, 40, 45 and 50 Tesla as a function
of temperature.  Parameters are as in Fig. 1, in particular, $T_{KT}=80$ K.} \label{fig:DHnonX}
\end{figure}

Finally, we record the salient results of our analysis that can be
tested against careful measurements:

\noindent i) A logarithmic variation of the magnetization with
field in the low-field two-dimensional vortex liquid below
$T_{KT}$, with a coefficient set by the superfluid density and the
temperature.
Specifically, \ba  \frac{\partial M}{\partial \log H} \approx
\frac{\pi \rho_s(T)}{ d \phi_0}\left(
1-\frac{2 k_B T}{\pi \rho_s(T)}\right).\nonumber
\ea
Thus the theory predicts this simple relation between these
directly measurable quantities, that can be checked in any
sufficiently two-dimensional material where this regime should
exist.

\noindent
ii) The double logarthmic variation of $M$ in Eq.~(\ref{eq:critmag})
 at $T_{KT}$ or at intermediate fields away from $T_{KT}$.
Optimistically, one might hope for a direct fit to this functional
form. This field dependence also implies that in an $M$ vs. $\log
H$ plot, the curves for $T<T_{KT}$, though straight in the
low-field regime, should exhibit an upward curvature at larger
fields (see Fig. \ref{fig:DHLog}).

\noindent iii) Correspondingly, plots of $M(T)$ at different
fields should exhibit a systematic drift of the ``crossing point".
This is clear from plotting our results, as in Fig.
\ref{fig:DHnonX}, which explicitly shows the downward creep of the
``crossing point" towards $T_{KT}$ from above as the field is
decreased within the two-dimensional vortex liquid regime.
This trend definitely appears to be there in the recent BSSCO data
of Wang, {\it et al.} \cite{WangM} for underdoped and optimally
doped samples.

\section{Summary and future directions}
The mechanism of superconductivity in the cuprates remains one of
the outstanding puzzles of the physics of correlated electrons.
Nevertheless the proposition that aspects of their
finite temperature behavior can be understood as consequences
of sizeable thermal fluctuations of the superconducting order
parameter has gained support in recent work.  In
this work we have examined the effects of such fluctuations in the
two dimensional limit near the Kosterlitz-Thouless transition and
presented an asymptotically exact calculation of the magnetization
in this vortex liquid state.
The preliminary comparison with the highly anisotropic cuprates such
as BSSCO is encouraging.

On the theoretical side, it would be very
useful to extend our calculation to higher fields, e.g., by keeping
terms of higher order in vortex density in deriving flow Eqs.~(\ref{eq:dy}), (\ref{eq:dx}) and (\ref{eq:dA}).
The high-field behavior is one aspect of experimental
magnetization data that does not appear to be captured well by our
calculation. Strong signatures of superconducting fluctuations are
also present in the Nernst coefficient. Indeed,
Ref.~\onlinecite{WangM} has reported that the magnetization and
the Nernst signal track each other. It would be desirable to have
a theory of the Nernst effect near the Kosterlitz-Thouless
transition. We hope to report on this in the near future.

\acknowledgements We are very grateful to L. Li, Y. Wang and N.P.
Ong for sharing their data with us. We would also like to thank I.
Ussishkin, S. Kivelson, S. Mukerjee and W. Wu for instructive
discussions and the National Science Foundation through its grant
NSF-DMR-0213706 (DAH, VO, SLS) and the David and Lucille Packard
Foundation (SLS, VO) for support.

\appendix
\section{Superconductor in a field to Coulomb plasma}
For completeness, we sketch the mapping between a superconductor in a
transverse field and a non-neutral Coulomb plasma\cite{Minnhagen}. The Hamiltonian
of a two-dimensional superconductor with a uniform magnetic
induction $\mu_0H_{c1}\ll B\ll \mu_0H_{c2}$ can be approximated as
\ba H_{\rm SC}=\int d^2 x \frac{\rho_{s0}}{2}|\nabla
\theta-\frac{2e}{\hbar c}{\bf A}|^2, \ea where $\theta, {\bf A}\
{\rm and}\ B=\hat{\bf z}\cdot\nabla \times {\bf A}$ are the usual
phase, gauge and perpendicular magnetic fields.  We also impose
the constraint that the net vortex charge density matches $B$.
 Next, we explicitly separate the phase field into its
longitudinal and transverse components by introducing a spin-wave
field $\phi$ and vortex charge density field $n$ and
Fourier-transform the Hamiltonian \ba
&&H_{\rm SC}=L^2\sum_{\bf q\neq 0} h_{\bf q} \\
&&h_{\bf q}=\frac{\rho_{s0}}{2}|{\bf q} \phi_{\bf q} +n_{\bf
q}\frac{{\bf z}\times
{\bf q}}{q^2}|^2 \\
&&=\frac{\rho_{s0}}{2}|{\bf q} \phi_{\bf q}|^2
+\frac{\rho_{s0}}{2}|n_{\bf q}\frac{{\bf z}\times {\bf
q}}{q^2}|^2. \ea  We then ignore the spin-wave part as it
decouples from the rest of the problem.
Transforming back into real space we arrive at the Hamiltonian of
a non-neutral two component Coulomb gas \ba H_{\rm CG}=\frac{\pi
\rho_{s0}}{4}\int d^2 {\bf x} d^2 {\bf y} \delta n({\bf x})
\log \left[\frac{({\bf x}-{\bf y})^2}{a_0^2}\right]\delta
n({\bf y}), \ea where $\delta n({\bf y})=n({\bf
y})-B/\phi_0$.  The range of integration above
excludes the infinite self-interaction ${\bf x=y}$ of each vortex.
One last remaining ingredient is the bare core energy $E_{c0}$
determined by the physics omitted in this derivation, it is
usually taken as the energy cost of suppressing the order
parameter inside the vortex core.  So, finally, the Hamiltonian of
a superconductor in the external field is written $H_{\rm
SC}=H_{\rm CG}+E_{c0} N_T$, which is a continuum version of Eq.~(7).

\end{document}